\documentclass[11pt,twoside]{article}

\usepackage{asp2006}
\usepackage{epsf}
\usepackage{psfig}
\usepackage{lscape}
\usepackage{natbib}
\bibpunct{(}{)}{;}{a}{}{,}

\begin{document}
\title{The Age Distributions of Clusters and Field Stars in the SMC --- Implications for Star Formation Histories}
\author{J.~M.~Diederik~Kruijssen$^\star$ and Henny~J.~G.~L.~M.~Lamers}
\affil{Astronomical Institute, Utrecht University, Princetonplein 5, NL-3854 CC Utrecht, The Netherlands. $^\star$E-mail: kruijssen@astro.uu.nl}

\begin{abstract}
Differences between the inferred star formation histories (SFHs) of star clusters and field stars seem to suggest distinct star formation processes for the two. The Small Magellanic Cloud (SMC) is an example of a galaxy where such a discrepancy is observed. We model the observed age distributions of the SMC clusters and field stars using a new population synthesis code, SPACE, that includes stellar evolution, infant mortality and cluster dissolution. We find that the two observed age distributions can be explained by a single SFH, thus eliminating the need to assume two separate mechanisms for star formation.
\end{abstract}

\section{Introduction}
We consider the relation between age distributions of clusters and field stars. It has been suggested that separate formation mechanisms for clusters and field stars exist in galaxies \citep[e.g.,][]{meurer95,chandar05}, implying different formation histories for the two. However, this appears to contradict theory and observations, as most stars are supposed to be formed in clusters.

Observations of the Small Magellanic Cloud (SMC) show a significant difference between the age distributions of clusters and field stars \citep{hodge87,chandar06}. The SMC therefore provides an ideal opportunity to investigate whether the observed distributions can be reproduced by one star formation history (SFH) if all stars were formed in clusters.

\section{Model Ingredients}
To reproduce the SMC cluster and field star age distributions, we first adopt the global SFH for the SMC derived from colour-magnitude diagrams by \citet{harris04}. Secondly, clusters are assumed to be initially distributed according to a cluster initial mass function (CIMF) represented by a power law with index -2. Of these clusters, a mass-independent fraction $f_{\rm IMR}$ (infant mortality rate) does not survive its first 10~Myr due to residual gas expulsion by supernovae and the consequent removal of binding energy. The surviving clusters gradually dissolve due to the tidal field and external perturbations. The loss of stars due to these effects is described by an analytical model \citep{lamers05,kruijssen08}. Dynamical cluster mass loss follows an exponential decrease on an instantaneous dissolution timescale related to present cluster mass as $\tau_{\rm dis}=t_0 {M_{\rm cl}^{\rm tot}}^\gamma$ with $\gamma=0.62$ and $t_0$ a parameter describing the speed of dissolution. We monitor the transition of stars from clusters into the field.

\section{Age distributions and implications for star formation histories}
\begin{figure}[!t]
\plotone{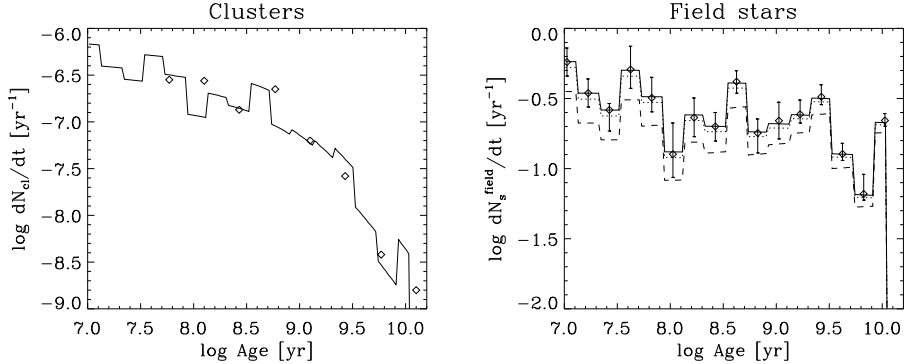}
\caption[]{\label{fig:kruijssen1}
      Age distributions of SMC clusters (left) and field stars (right). Diamonds denote observed values \citep{hodge87,chandar06}, while our model predictions are shown for infant mortality rates of $f_{\rm IMR}=\{0.6, 0.9, 0.995\}$  as dashed, dotted and solid lines, respectively. Error bars represent 2-$\sigma$ error margins.
    }
\end{figure}
The observed age distributions and the fit from the cluster population synthesis model SPACE \citep{kruijssen08b} are shown in Figure~\ref{fig:kruijssen1}. Assuming the SFH from \citet{harris04}, we see that the shape of the cluster age distribution is accurately described by adopting a dissolution timescale of $t_0=30$~Myr (implying a total disruption time of $\sim 8.5$~Gyr for a $10^4$ $M_{\odot}$ cluster).

In principle, it is possible to derive the infant mortality rate from the ratio of the number of clusters and field stars. However, only part of the SMC is covered in the cluster sample from \citet{hodge87}. Therefore, only an upper limit for the infant mortality rate can be derived from the available data. We find this upper limit to be $f_{\rm IMR}=0.995$. Figure~\ref{fig:kruijssen1} (right panel) shows that the field star age distribution is still well-fitted by a more reasonable value of $f_{\rm IMR}=0.9$. This implies that it is not significantly affected by the partial cluster coverage of the SMC. The overpopulation of field stars relative to clusters at old ages is caused by cluster dissolution. We conclude that {\it no such process as separate field star formation is required to explain the age distributions of field stars and clusters}.

\acknowledgements
JMDK is grateful to the Leids Kerkhoven-Bosscha Fonds for supporting attendance of the conference.

\bibliographystyle{aspbib}
\bibliography{mybib}

\end{document}